\documentclass[aps,prl,twocolumn,groupedaddress,showpacs]{revtex4}
\usepackage{graphicx}

\begin{document}


\title{Mapping the magneto-structural quantum phases of Mn$_3$O$_4$}



\author{M. Kim, X.M. Chen, E. Fradkin, P. Abbamonte, S.L. Cooper}

\affiliation{
Department of Physics and Frederick Seitz Materials Research Laboratory, University of Illinois, Urbana, Illinois  61801, USA
}

\date{\today}

\begin{abstract}

We present temperature-dependent x-ray diffraction and temperature- and field-dependent Raman scattering studies of single crystal Mn$_3$O$_4$, which reveal the novel magnetostructural phases that evolve in the spinels due to the interplay between strong spin-orbital coupling, geometric frustration, and applied magnetic field. We observe a structural transition from tetragonal to monoclinic structures at the commensurate magnetic transition at T$_2$=33K, show that the onset and nature of this structural transition can be controlled with an applied magnetic field, and find evidence for a field-tuned quantum phase transition to a tetragonal incommensurate or spin glass phase.

\end{abstract}

\pacs{71.70.Ej, 73.43.Nq, 78.30.-j}

\maketitle


Strong coupling among the spin, lattice, and orbital degrees of freedom in the geometrically frustrated spinel compounds [1] results in a rich variety of exotic magnetic and structural phases and properties that are of both scientific and technological interest.  For example, the chromium-oxide spinels AB$_2$O$_4$ (A=Zn, Cd, Hg; B=Cr) exhibit three-dimensional spin-Peierls transitions involving coupled magnetic and structural transitions;[2-6] the vanadium-oxide spinels AB$_2$O$_4$ (A=Zn, Cd, Mn; B=V) display complex spin/orbital ordering and technologically useful phenomena such as large magnetoelastic and magnetodielectric effects;[7-14] and sulfer-based spinels such as FeB$_2$S$_4$ (B=Cr,Sc) [15] exhibit orbital-liquid or -glass ground states in which frustration prevents orbital ordering down to T=0.  The rich magneto-structural phases of the spinels are thought to be governed by the interplay between spin-orbital coupling, applied magnetic field, and frustrated exchange interactions,[1,16] but there has been little experimental investigation of the microscopic details of this interplay.

The binary spinel Mn$_3$O$_4$ is a relatively simple system for experimentally studying the complex interplay between structure, spin-orbital coupling, and magnetic field in the spinels:  in spite of its simpler chemical composition---with Mn ions at both tetrahedral (A=Mn$^{2+}$) and octahedral (B=Mn$^{3+}$) sites---Mn$_3$O$_4$ exhibits the rich magneto-structural transitions characteristic of more complex ternary magnetic spinels:  Below T$_C$=43K, the spins in Mn$_3$O$_4$ exhibit Yafet-Kittel-type ferrimagnetic ordering, in which the net spin of the octahedrally coordinated Mn$^{3+}$ spins is antiparallel to the [110] direction of the tetrahedrally coordinated Mn$^{2+}$ spins, with pairs of Mn$^{3+}$ spins canted by ${\pm}{\theta}_{YK}$ from the [$\overline{1}\hspace{1pt}\overline{1}0$] direction, where ${\cos}{\theta}_{YK}$=0.38 (0.33) at T=4.7K and ${\cos}{\theta}_{YK}$=0.40 (0.25) at T=29K for the ``non-doubling'' octahedral (``doubling'' octahedral) site.[14,17]  However, below T$_1$=39K, Mn$_3$O$_4$ develops an incommensurate sinusoidal or spiral spin structure of the Mn$^{3+}$ spins; and below T$_2$=33K, Mn$_3$O$_4$ exhibits a commensurate spin structure in which the magnetic unit cell doubles the chemical unit cell.[17,18]  Recent studies also show that the magnetic transitions in Mn$_3$O$_4$ are are associated with significant temperature- and field-dependent changes in the dielectric constant and lattice parameters, reflecting strong spin-lattice coupling in this material.[19,20]

In this paper, we report Raman spectroscopy and x-ray diffraction measurements of the temperature- and magnetic-field-dependent phases of single-crystal Mn$_3$O$_4$.  These combined measurements offer a particularly clear, microscopic view of the diverse magneto-structural phases that can result from the interplay between strong spin-orbital coupling, geometric frustration, and applied magnetic field in the spinels; this includes evidence for a quantum phase transition to a tetragonal spin/orbital glass phase for intermediate fields with H$\parallel$[$1\overline{1}0$], which we propose is caused by a field-tuned degeneracy between magneto-structural states.

A single-crystal sample of Mn$_3$O$_4$ was grown at the University of Illinois using a floating zone technique; the sample was identified as a single phase crystal using both x-ray powder diffraction with a Rigaku D-Max system and a pole figure analysis with a Phillips X'pert system.  Field-dependent Raman measurements were performed as described previously [21] on an as-grown surface of single crystal Mn$_3$O$_4$ having a surface normal along the [110] direction.  Temperature-dependent x-ray measurements were carried out using a Rigaku rotaflex RU-300 with a closed-cycle He refrigerator in the range of 10K to 75K. In addition, a Phillips MRD X'Pert was used for high precision measurements at room temperature. Both measurements were performed on the as-grown [110] surface of single crystal Mn$_3$O$_4$. A least squares program was used to determine the lattice parameters of the crystal from the data.

The room-temperature Raman spectrum of our Mn$_3$O$_4$ sample exhibits 5 phonon peaks, consistent with previous reports:[22] a $T_{2g}$ symmetry mode at 290 cm$^{-1}$, an $E_g$ symmetry mode at 320 cm$^{-1}$, $T_{2g}$ symmetry modes at 375 cm$^{-1}$ and 479 cm$^{-1}$, and an $A_{1g}$ symmetry mode at 660 cm$^{-1}$.[22,23]  In this paper, we focus on the lowest energy $T_{2g}$ phonon mode, which is associated with Mn-O bond-stretching vibrations of the tetrahedral site ions.[22,23]

\begin{figure}[tp]
\centering
\includegraphics[width=8cm]{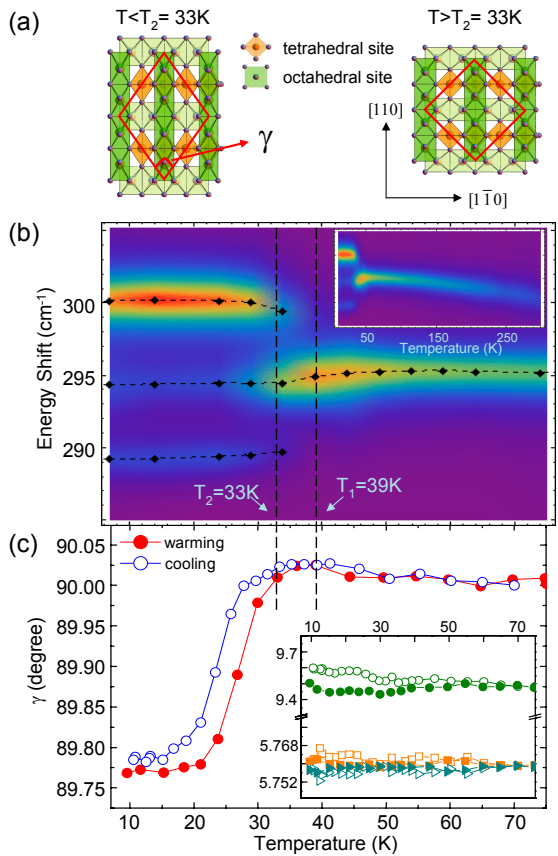}
\vspace{-7pt}
\caption{(a) Illustrations of the monoclinic structure for T$<$T$_2$=33K and the tetragonal structure of Mn$_3$O$_4$ for T$>$T$_2$=33K. (b) Contour plot of the $T_{2g}$ phonon mode intensity as functions of energy and increasing temperature, where red=700 counts and blue=0 counts. (inset) Contour plot of the $T_{2g}$ phonon mode intensity over the full temperature range 7-290K. (c) Temperature dependence of ${\gamma}$---the angle between the $a$- and $b$-axis directions---as functions of increasing temperature (closed symbols) and decreasing temperature (open symbols). (inset) Temperature dependence [in K] of lattice constants $a$ (squares), $b$ (triangles), and $c$ (circles) [in {\AA}] for Mn$_3$O$_4$.  Open (closed) symbols represent measurements taken with decreasing (increasing) temperature.} \label{figure1}
\vspace{-17pt}
\end{figure}

Fig. 1(b) shows the temperature dependence of the $T_{2g}$ mode intensity and the $T_{2g}$ mode energy and linewidth for light polarized along the [$1\overline{1}0$] crystallographic direction of Mn$_3$O$_4$.  Three distinct temperature regimes can be identified:  (i) T $>$ T$_1$=39K -- Above T$_1$=39K, the $T_{2g}$ mode narrows and shifts to higher frequencies with decreasing temperature in a conventional manner, i.e., consistent with a temperature dependence governed by anharmonic (multi-phonon) effects.[24]  (ii) T$_2$=33K $<$ T ${\leq}$ T$_1$=39K -- In the incommensurate magnetic phase regime between T$_2$=33 K and T$_1$=39K, the lowest $T_{2g}$ mode decreases in energy slightly with decreasing temperature due to magnetoelastic effects, but exhibits no evidence for a change in structural symmetry.  (iii) T $<$ T$_2$=33K -- Below the commensurate magnetic transition T$_2$=33K, the $T_{2g}$ mode abruptly splits into three modes near 290 cm$^{-1}$, 295 cm$^{-1}$, and 300 cm$^{-1}$.  This splitting is consistent with a tetragonal-to-monoclinic distortion below T$_2$, which splits the degenerate $T_{2g}$ mode by expanding the Mn$^{2+}$-O$^{2-}$ bond length along the easy-axis [110] direction and contracting the Mn$^{2+}$-O$^{2-}$ bond length along the hard-axis [$1\overline{1}0$] direction (see illustrations, Fig. 1(a)).  Note that the relative intensities of the three modes shown in Fig. 1(b) for T $<$ T$_2$ confirm that the Mn$^{2+}$-O$^{2-}$ bond length expands along the easy-axis [110] direction below T$_2$=33K:  the higher-energy ($\sim$300 cm$^{-1}$) split mode---which is associated with vibrations of the contracted Mn$^{2+}$-O$^{2-}$ bond---exhibits the strongest light scattering intensity, indicating that the contracted Mn$^{2+}$-O$^{2-}$ bond is oriented in the direction of the incident light polarization, i.e., along the [$1\overline{1}0$] direction.  By contrast, the intensity of the lower-energy ($\sim$290 cm$^{-1}$) split mode---which is associated with vibrations of the expanded Mn$^{2+}$-O$^{2-}$ bond---has a substantially weaker intensity than the $\sim$300 cm$^{-1}$ mode, consistent with an expansion of the Mn$^{2+}$-O$^{2-}$ bond in a direction perpendicular to the incident light polarization, i.e., along the [110] direction.

To provide more definitive evidence for a tetragonal-to-monoclinic phase transition below T$_2$ in Mn$_3$O$_4$, temperature-dependent x-ray diffraction measurements were performed on the same crystal.  Fig. 1(c) shows the temperature dependence of the lattice parameters $a$, $b$, $c$, and $\gamma$---the angle between $a$ and $b$---as functions of both increasing (``warming'') and decreasing (``cooling'') temperature.  While $\gamma$ exhibits an abrupt decrease near T$_2$=33K---indicating an abrupt decrease in the angle between $a$- and $b$-axis directions at this temperature, the lattice parameters $a$ and $b$ exhibit no significant temperature dependence, and the lattice parameter $c$ exhibits only a weak temperature dependence and hysteretic behavior.  This behavior confirms that Mn$_3$O$_4$ exhibits a first-order tetragonal-to-monoclinic structural phase transition near T$_2$=33K, as illustrated schematically in Fig. 1(a).
Notably, the observed monoclinic crystal structure we observe below T$_2$=33K is consistent with the spin structure that has been previously reported for Mn$_3$O$_4$ below T$_2$:  the coplanar spin structure of the doubled unit cell has spins lying within the ($1\overline{1}0$) plane with a net spin along [110], in which the Mn$^{3+}$ spins are canted by an angle ${\pm}{\theta}_{YK}$ from the [$\overline{1}\hspace{2pt}\overline{1}0$] direction.  This spin canting is associated with a tilting of the $d_{3z^2-r^2}$  orbitals of Mn$^{3+}$ toward the [$\overline{1}\hspace{1pt}\overline{1}0$] direction due to spin-orbital coupling, which results in a tilting of Mn$^{3+}$ octahedra and an expansion of the Mn$^{2+}$-O$^{2-}$ bond length along the [110] direction.[17,20]  We also note that the tetragonal structure we observe for the incommensurate magnetic phase regime between T$_2$=33K and T$_1$=39K (Fig. 1(b)) is consistent with the presence of an axially symmetric spiral spin structure [17] in this temperature regime.

\begin{figure}[tp]
\centering
\vspace{-2pt}
\includegraphics[width=8cm]{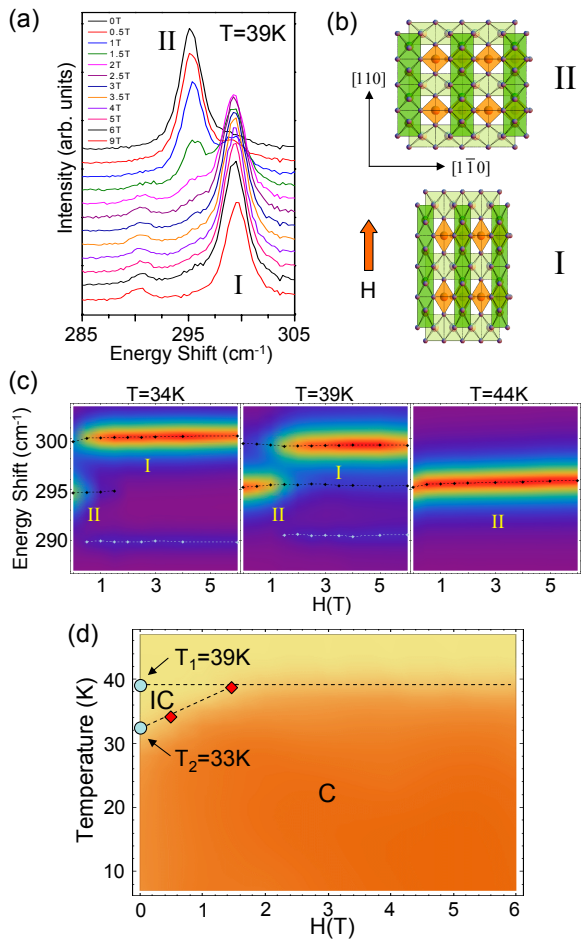}
\vspace{-7pt}
\caption{(a) Field dependence of the Raman spectra at T=39K for H$\parallel$[110].  (b) Illustrations of the Mn$_3$O$_4$ structure in (top) the low-field undistorted phase and (bottom) the high-field monoclinic phase. (c) Contour plots of the intensities of the split modes as functions of energy and field at (left) T=34K, (middle) 39K, and (right) 44K, where red=700 counts and blue=0 counts for H$\parallel$[110]. (d) Phase diagram as functions of magnetic field (along [110]) and temperature; orange region=structure I, yellow region=structure II, IC=incommensurate magnetic phase, C=commensurate (cell-doubled) magnetic phase.} \label{figure2}
\vspace{-15pt}
\end{figure}

\begin{figure}[tp]
\centering
\vspace{-2pt}
\includegraphics[width=8cm]{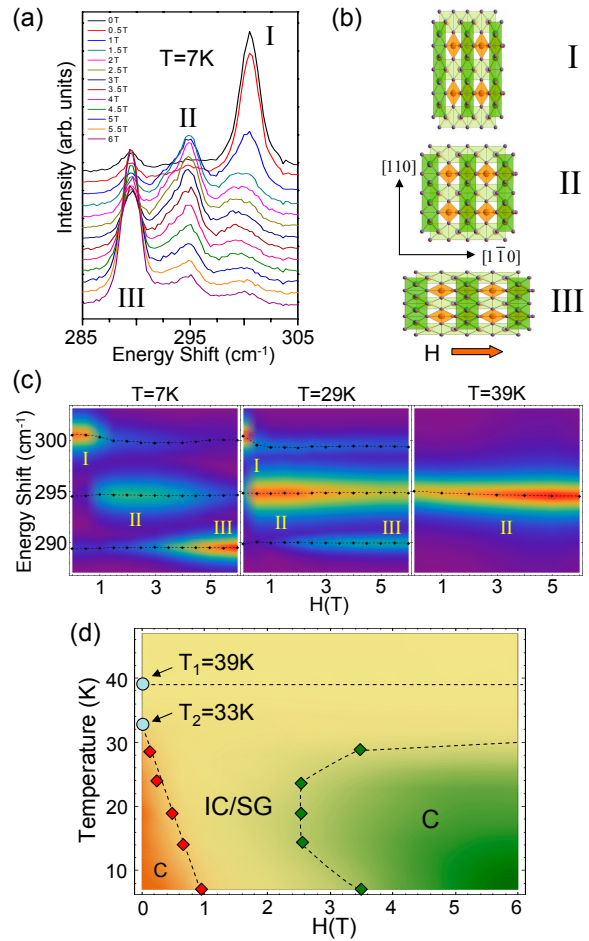}
\vspace{-7pt}
\caption{(a) Field dependence of the Raman spectra at T=7K for a magnetic field applied along the hard-axis [$1\overline{1}0$] direction.  (b) Illustrations of the Mn$_3$O$_4$ structure in (top) the low-field monoclinic phase, (middle) the intermediate undistorted tetragonal phase, and (bottom) the high-field monoclinic phase. (c) Contour plots of the intensities of the split modes as functions of energy and field at (left) T=7K, (middle) 29K, and (right) 39K, where red=700 counts and blue=0 counts for H$\parallel$[$1\overline{1}0$]. (d) Phase diagram as functions of magnetic field (along [$1\overline{1}0$]) and temperature; orange region=structure I, yellow region=structure II, green region=structure III, IC/SG=incommensurate or spin glass magnetic phase, and C=commensurate (cell-doubled) magnetic phase.} \label{figure3}
\vspace{-15pt}
\end{figure}

The distinctive Raman spectroscopic signatures associated with the different magneto-structural phases in Mn$_3$O$_4$---illustrated in Fig. 1(b)---offer a convenient method for investigating the phases induced with an applied magnetic field.  Fig. 2 illustrates the field-induced structural phases of Mn$_3$O$_4$ for fields applied along the easy-axis [110] direction.  Fig. 2(a) and the left and middle plots of Fig. 2(c) show that, in the incommensurate magnetic phase regime T$_2$=33K $<$ T ${\leq}$ T$_1$=39K, the ${\sim}$295cm$^{-1}$ mode associated with the undistorted tetragonal phase (structure II in Fig. 2(b)) exhibits a field-induced splitting similar to that induced upon cooling below T$_2$=33K in zero magnetic field (Fig. 1(b)).  Thus, in the incommensurate magnetic phase, an applied magnetic field along the easy axis [110] direction induces a tetragonal-to-monoclinic distortion in Mn$_3$O$_4$ by forcing the Mn$^{3+}$ spins to order within the ($1\overline{1}0$) plane and inducing the cell-doubled coplanar magnetic structure associated with the monoclinic structure (see Fig. 2(b)).  This strong magnetoelastic response at rather modest fields likely arises from a field-induced increase---via spin-orbit coupling---in the hybridization between the $d_{3z^2-r^2}$ and $d_{xy}$ orbitals of Mn$^{3+}$ for H$\parallel$[110], and demonstrates that the magnetostructural states I and II in Fig. 2(b) have very similar free energies.  Fig. 2(d) summarizes the different magnetic/structural phases of Mn$_3$O$_4$ as functions of magnetic field and temperature for H$\parallel$[110].

A richer magneto-structural phase diagram is observed when the magnetic field is applied along the hard-axis [$1\overline{1}0$] direction of Mn$_3$O$_4$, as illustrated in Fig. 3.  Figs. 3(a) and 3(c) show that three different phase regimes are apparent for H$\parallel$[$1\overline{1}0$] at T=7 K.  For H $<$ 1 T, the high-energy 300cm$^{-1}$ mode is the dominant mode, indicating that the monoclinic distortion with an expanded Mn$^{2+}$-O$^{2-}$ bond length along the easy-axis [110] direction (structure I) persists at low fields.  On the other hand, at high magnetic fields, i.e., for H $>$ 4T for T=7K, Figs. 3(a) and (c) show that the low-energy 290cm$^{-1}$ mode is most intense, indicating a monoclinic phase in which the field reorients the Mn spin---and expands the Mn$^{2+}$-O$^{2-}$ bond lengths---along the hard-axis [$1\overline{1}0$] direction (structure III).

Most remarkably, however, Figs. 3(a) and 3(c) show that the field-induced transition from a monoclinic distortion with M$\parallel$[110] to a monoclinic distortion with M$\parallel$[$1\overline{1}0$] is not abrupt, but occurs via an intermediate field regime (1T $<$ H $<$ 4T) in which the dominant mode is the $\sim$295cm$^{-1}$ mode, i.e., the mode associated with the undistorted tetragonal phase (structure II in Figs. 3(b) and 3(c)) observed in the paramagnetic and incommensurate phases above T$_2$=33K for H=0 (Fig. 1(b)).  We suggest that the quantum phase transition to this intermediate phase is a transition from a ferrimagnetic, monoclinic phase with M$\parallel$[110], to a ``spin/orbital'' glass phase---in which the Mn spins are randomly oriented along the [110] and [$1\overline{1}0$] directions---or to an incommensurate spiral spin phase; both of these possibilities are consistent with a tetragonal structure (see Fig. 1(b)).  Fig. 3(c) shows that, with increasing temperatures, this field-induced tetragonal regime becomes more pronounced.  Fig. 3(d) summarizes the different magnetic/structural phases of Mn$_3$O$_4$ as functions of magnetic field and temperature for H$\parallel$[$1\overline{1}0$].

The results summarized in Fig. 3(d) indicate that the competition between spin-orbital coupling, geometrical frustration, and applied magnetic field in Mn$_3$O$_4$ leads to an incommensurate---or even glassy---spin state at T=0 that is sandwiched---as a function of applied field with H$\parallel$[$1\overline{1}0$]---between commensurate spin phases.  The observation of a quantum phase transition to an intermediate incommensurate or spin-glass tetragonal phase (structure II in Fig. 3(b))---rather than a simple metamagnetic transition between or coexistence of commensurate ferrimagnetic phases (structures I and III in Fig. 3(b)) at intermediate fields---likely reflects the importance of spin-lattice coupling in Mn$_3$O$_4$, wherein the balancing of elastic and magnetic energies favors the formation of a more isotropic magnetostructural configuration at intermediate fields with H$\parallel$[$1\overline{1}0$].  Indeed, this more isotropic magnetostructural configuration (II) appears to be the means by which Mn$_3$O$_4$ resolves the frustration that arises from the field-induced degeneracy between M$\parallel$[110] and M$\parallel$[$1\overline{1}0$] spin configurations at intermediate field values with H$\parallel$[$1\overline{1}0$].  Field-dependent neutron scattering and heat capacity measurements would be useful for clarifying the nature of the spin configuration and entropy of this highly frustrated field-induced state.  It is also important to study the T$\sim$0 spin dynamics of this intermediate-field isotropic phase, in particular to explore the extent to which quantum critical fluctuations govern the dynamics in this frustrated phase regime.

In summary, combined temperature- and field-dependent Raman scattering and temperature-dependent x-ray scattering studies provide a clear microscopic view of the diverse and complex magnetostructural phases that evolve in the spinel material Mn$_3$O$_4$ due to the interplay between spin-orbital coupling, geometrical frustration, and applied magnetic field.  In addition to identifying the specific structural phases associated with the different magnetic states observed in Mn$_3$O$_4$ with H=0, we have identified the microscopic magnetostructural changes that are associated with the novel magnetodielectric behavior previously observed for this material.  Most interesting is the observation of a quantum phase transition to a structurally isotropic, incommensurate/disordered spin state for intermediate fields with H$\parallel$[$1\overline{1}0$], which reflects a compromise this system takes to accommodate the frustration imposed by a field-induced degeneracy between differing magnetostructural states.

This material is based on work supported by the U.S. Department of Energy, Division of Materials Sciences, under Award No. DE-FG02-07ER46453, through the Frederick Seitz Materials Research Laboratory at the University of Illinois at Urbana-Champaign, and by the National Science Foundation under Grant NSF DMR 08-56321.



\begin{thebibliography}{}
\bibitem{23} Following reference 22, we use the cubic space group notation to index the Raman-active phonons in tetragonal Mn$_3$O$_4$ because the Raman spectrum of Mn$_3$O$_4$ corresponds well to those of cubic spinels.
\end{thebibliography}
\end{document}